УДК 004.89:004.424.4:519.87

# Реализация генетического алгоритма для эффективного документального тематического поиска

**Иванов Владимир Константинович**, к.т.н., доцент (Тверской государственный технический университет, директор Центра научно-образовательных электронных ресурсов, наб. Аф. Никитина, 22, г. Тверь, 170026, Россия, mtivk@mail.ru, тел. +79051272231, почтовый адрес автора: просп. 50-лет Октября, д. 3, корп. 1, кв. 142, г. Тверь, 170024).

**Мескин Павел Иосифович** (Тверской государственный технический университет, ведущий программист Центра научно-образовательных электронных ресурсов, наб. Аф. Никитина, 22, г. Тверь, 170026, Россия, pavel.meskin@gmail.com, тел. +79206928127, почтовый адрес автора: ул. Советская, д. 60, кв. 36г. Тверь, 170100).

**Аннотация**

Качество документального тематического поиска, то есть поиска документов, содержащих координированную информацию в заданном тематическом сегменте, не всегда удовлетворительно. Несмотря на наличие мощных поисковых систем для информационных ресурсов Интернет или для специализированных баз данных, процесс продолжает оставаться трудоемким и слабо поддержанным программно и методологически.

В настоящей статье описывается программная реализация генетического алгоритма для выявления и отбора наиболее релевантных результатов, полученных в ходе последовательно выполняемых операций тематического поиска. При этом моделируется эволюционный процесс, который формирует устойчивую и эффективную популяцию поисковых запросов, образует поисковый образ документов или семантическое ядро, создает релевантные множества искомых документов, позволяет автоматически классифицировать результаты поиска. В статье обсуждаются особенности тематического поиска, обосновывается применение генетического алгоритма, описываются аргументы целевой функции, рассматриваются основные шаги и параметры алгоритма. Отмечается, что целевая функция или критерий качества поиска определяется позицией документа в списках результатов, построенных поисковой системой при выполнении максимального числа различных запросов, и семантической близостью к поисковому образу документов заданной тематики. Достаточно подробно описана программная реализация: основные объектные модели, пользовательский интерфейс, основная библиотека алгоритма, модули морфологического анализа, семантического анализа сходства текстов, поиска, управления базой данных, управления метаданными. Приводятся сведения о составе классов модулей и их компонентов.

В заключении отмечается, что реализованный генетический алгоритм, является одним из элементов программного обеспечения разрабатываемой интеллектуальной системы информационной поддержки инноваций в науке и образовании. Он играет важную роль в обеспечении адаптивности функционирования поисковых механизмов. А разработанное программное обеспечение алгоритма создает достаточно широкий базис для дальнейших исследований и разработок.

**Ключевые слова**: генетический алгоритм, документ, объектная модель, мутация, поисковый запрос, популяция, приспособленность, ранжирование, реализация программного обеспечения, релевантность, скрещивание, тематический поиск, фильтрация.

UDC 004.89:004.424.4:519.87

# Implementation of Genetic Algorithm for Effective Document Subject Search

**Ivanov V.K.**, Ph.D., Associate Professor (Tver State Technical University, 22, Quay Nikitin, Tver, 170026, Russian Federation, mtivk@mail.ru)
**Meskin P.I.**, Lead Programmer (Tver State Technical University, 22, Quay Nikitin, Tver, 170026, Russian Federation, pavel.meskin@gmail.com)

**Abstract**

The quality of documentary subject search or search for documents containing specifically coordinated information on a target subject is not always satisfactory. Despite the availability of powerful search engines for information resources on the Internet or special databases, the process remains time-consuming and poorly supported by software and methodologically.

This paper describes the software implementation of genetic algorithm for identifying and selecting most relevant results received during sequentially executed subject search operations. Simulated evolutionary process generates sustainable and effective population of search queries, forms search pattern of documents or semantic core, creates relevant sets of required documents, allows automatic classification of search results. The paper discusses the features of subject search, justifies the use of a genetic algorithm, describes arguments of the fitness function and describes basic steps and parameters of the algorithm. It is noted that the fitness function or quality criteria determined by the position of the document in search results built by the search engine for maximum number of different queries and semantic similarity of search pattern of documents on a given subject. Software implementation is described in detail: the general object model, user interface, the main library of the algorithm, morphological analysis module, texts similarity analysis module, search module, database management module, metadata management module. The information on module classes composition and components is provided.

The paper describes genetic algorithm software implementation that is one of the elements of project Intelligent Distributed Information Management System for Innovations in Science and Education powered by the Russian Foundation of Basic Research. The algorithm plays an important role in functioning of the adaptive search engines. It is noted that developed algorithm software creates a sufficiently broad basis for further research and development.

**Введение**

Качество документального тематического поиска, то есть поиска документов, содержащих координированную информацию в заданном тематическом сегменте, не всегда удовлетворительно. Несмотря на наличие мощных поисковых систем для информационных ресурсов Интернет или для специализированных баз данных, процесс продолжает оставаться трудоемким и слабо поддержанным программно и методологически.

В настоящей статье описывается реализация генетического алгоритма для выявления и отбора наиболее релевантных результатов, полученных в ходе последовательно выполняемых операций тематического поиска. Обсуждаются особенности тематического поиска, обосновывается применения генетического алгоритма, описываются компоненты целевой функции, рассматриваются основные шаги и параметры алгоритма. Описывается программная реализация: основные объектные модели, пользовательский интерфейс, основная библиотека алгоритма, модули морфологического анализа, семантического анализа сходства текстов, поиска, управления базой данных, управления метаданными.

Представляется, что описанные в настоящей статье подходы к организации тематического поиска могут быть успешно применены во многих областях. Например, обзоры источников научно-технической, коммерческой и социальной информации, поиск коммерчески ценной информации, сбор информации о клиентах, определение новых областей при бизнес-планировании, конкурентный анализ и разведка, поиск инновационных решений, подбор и экспертиза учебно-методических материалов, анализ конкурсной документации и условий экспертизы, экспертиза проектов, подборка материалов для патентных исследований.

**Особенности тематического поиска**

Поисковые системы Интернет обладают мощными механизмами быстрого и, во многих случаях, качественного поиска необходимой информации. Каждый пользователь Интернет искал в сети какие-либо конкретные факты и, как правило, находил либо точно то, что искал, либо что-то близкое. Поиск фактов – это поиск информационных объектов с определенными смысловыми и/или технологическими характеристиками. Результат такого поиска – описание объекта, события или явления с заданными значениями их свойств.

Другой вид поиска - тематический поиск. Здесь мы ищем целые категории (виды, роды) координированной информации в заданном тематическом сегменте, а не отдельные информационные объекты с заданными характеристиками (представителей этих категорий). Результатом тематического поиска помимо набора фактов следует считать сведения о ретроспективе, перспективе, взаимосвязях найденных информационных объектов, о текущих и вероятных трендах. Примеры тематического поиска: подборка материалов для патентных исследований или для обзоров источников научно-технической, коммерческой и социальной информации, поиск коммерчески ценных данных и т.п.

При выполнении тематического поиска неизбежно возникает ряд вопросов. Как совместно оценить релевантность документов, найденных разными запросами? Является ли ранжирование результатов поисковой системой корректным с позиций ожиданий пользователя? Все ли результаты, доступные для непосредственной оценки, соответствуют ожиданиям пользователя? Все ли результаты, соответствующие ожиданиям пользователя, попали в число доступных для непосредственной оценки? Как отфильтровать документы, не относящиеся по сути к искомой тематике? Могут ли быть обнаружены эффективные решения, которые относятся к другим областям применения, но могут быть успешно использованы в данной области? Невозможно однозначно ответить на эти вопросы в рамках тривиальных решений.

**Почему генетический алгоритм?**

Сложность формулировки точных запросов при тематическом поиске документов очевидна, и на это есть причины. Во-первых, искомая информация часто находится на стыке смежных областей. Во-вторых, одновременно с информацией о собственно предмете поиска (например, инновациях) желательно получать сведения о применениях, рисках, особенностях, пользователях, авторах, правообладателях, производителях. В-третьих, обычной является необходимость одновременного использования различных (иногда альтернативных) критериев отбора наиболее эффективных практик.

Как следствие, пользователи в поисковых запросах вынуждены применять множество сочетаний ключевых понятий, уточняя их в ходе анализа промежуточных результатов поиска. В результате в распоряжении будет большой объем результатов поиска (тысячи документов), в той или иной степени релевантных сформулированным запросам. При этом подробное рассмотрение всех найденных документов не производится (в большинстве случаев просматривается не более 2-3-х страниц результатов поиска).

Порядок действий пользователей при выполнении ими тематического поиска во многом напоминает эволюционный процесс. Пользователи ищут эффективные наборы и сочетания ключевых слов в запросах, имея в виду получение максимально релевантных результатов и параллельно анализируя содержание найденных материалов. Однако неочевидно, что при этом будет использована какая-либо обоснованная методика. Обычными подходами являются использование пользователями собственного опыта работы с материалами заданной тематики и/или уточнение запросов ключевыми понятиями из уже найденных пертинентных текстов.

Представляется, что для решения задач обработки результатов тематического поиска документов целесообразно использовать генетический алгоритм [1, 2 и др.], который может быть предназначен для организации эволюционного процесса, который:

а) формирует устойчивую и эффективную популяцию поисковых запросов;

б) образует соответствующий поисковый образ документов или семантического ядра;

в) приводит к получению релевантного искомого множества искомых документов;

г) создает предпосылки для автоматического получения классифицированного и ассоциированного множества документов.

**Целевая функция**

Исходная позиция для определения целевой функции предусматривает, что множество эффективных результатов поиска должно формироваться документами, которые удовлетворяют следующим условиям:

1. Документы находятся в первых позициях ранжированного списка результатов поиска, построенного поисковой системой.

2. Документы присутствуют в списках результатов, полученных при выполнении как можно большего числа различных запросов.

3. Документы семантически близки к поисковому образу (набору ключевых понятий) документов заданной тематики, в том числе эталонным текстам, формируемым при эволюции запросов.

Исходя из сказанного, для каждого $i$–го результата запроса целевая функция может быть определена как $w_i = f(p, r, s)$, где

$p$ – средний номер позиции документа в списке первых $P$ результатов выполненных поисковых запросов (учитываются только те списки результатов, где данный документ присутствует).

$r$ – количество появлений документа в результатах выполнения $N$ поисковых запросов. Отметим, что $r \leq N$ и $r = N,$ если документ появился в результатах выполнения всех запросов.

$s$ – семантическая близость текста найденного документа (или, по крайней мере, заголовка и сниппета - небольшого отрывка текста, используемого в качестве описания документа) и множества ключевых слов, формирующего поисковый образ документов заданной тематики.

Значение $w_i$ определяет ранг результата запроса. Значение целевой функции для каждого запроса вычисляется как средний ранг результатов этого запроса, а значение целевой функции для популяции запросов вычисляется как средний ранг запросов этой популяции.

В таком виде целевая функция используется в текущей реализации алгоритма.

**Описание алгоритма**

На рис. 1 приведена схема разработанного генетического алгоритма формирования эффективного множества запросов (алгоритма GAF). Ниже кратко описаны основные шаги алгоритма:

1. Подготовка ключевых слов, формирующих поисковый образ множества документов заданной тематики.

2. Выбор поисковой системы. Текущая реализация алгоритма предусматривает выбор поисковых систем Bing или Google, а также использование XQuery для поиска в базе данных XML-документов.

3. Формирование исходной популяции запросов (хромосом или особей) - комбинации ключевых понятий (генов) из поискового образа документов.

4. Выполнение запросов популяции – формирование множества дескрипторов найденных документов (заголовок, описание, адрес текста).

5. Вычисление значений целевой функции на основе веса каждого результата поиска $w_i = f_5 * p + f_6 * r + f_7 * s$, где $f_5, f_6, f_7$ – весовые коэффициенты, являющиеся параметрами алгоритма.

6. Селекция лучших запросов по их пригодности или значению целевой функции.

7. Выбор родительских пар запросов для формирования следующего поколения (популяции) запросов. Использован генотипный аутбридинг.

8. Скрещивание запросов. Операция реализуется дискретной рекомбинацией или одноточечным кроссинговером. Особенностью является генерация запросов-потомков с использованием синонимов для одновременного расширения базы ключевых понятий и наследования семантики запросов-родителей.

9. Вероятностная мутация запросов.

10. Формирование новой популяции – элитарный отбор из объединенной популяции запросов-родителей и запросов-потомков.

11. Остановка алгоритма – достижение стабильности популяции запросов или заданного (предельного) числа проходов алгоритма.

Отметим, что более подробно эти шаги описаны в [3]. Общая архитектура приложения, реализующего генетический алгоритм, представлена на рис. 2. Приложение построено на платформе .Net Framework. Основные функциональные компоненты следующие:

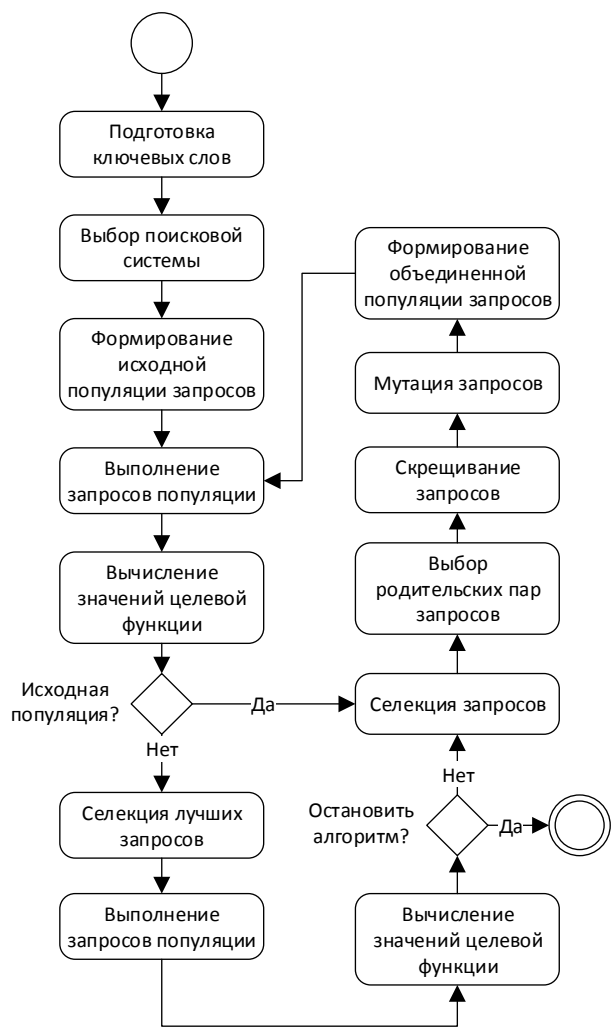
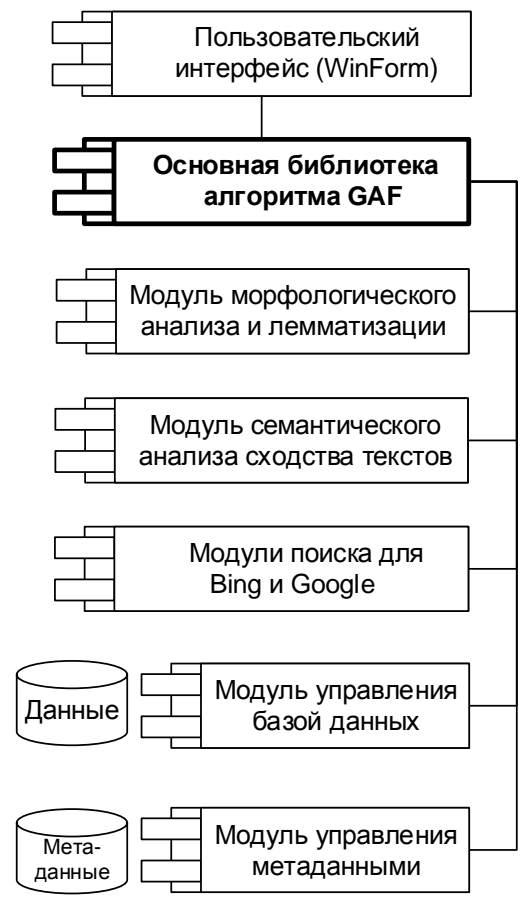

Рис.1. Схема генетического алгоритма GAF для формирования эффективного множества запросов

Рис. 2. Основные компоненты приложения, реализующего генетический алгоритм GAF

**Основная библиотека классов алгоритма GAF**

Объектная модель основной библиотеки классов алгоритма GAF в пространстве имен *rdomGAF* реализует классы, описание которых вместе с основными членами представлено в табл. 1. На рис. 3 приведено графическое изображение этой объектной модели.

Таблица 1

| Классы и их члены | Описание |
|---|---|
| Класс *GAF* - свойства, операции, входные и выходные данные генетического алгоритма. | |
| SetKeyWords() | Формирует список ключевых слов для запросов. |
| Stop() | Завершает алгоритм. |
| SelectBestIndividuals() | Отбирает лучшие запросы в популяции. |
| JoinPopulations() | Создает объединенную популяцию родителей и потомков. |
| CreateInitialPopulation() | Создает начальную популяцию запросов. |
| GetQueryResults() | Выполняет запрос в заданной поисковой системе и возвращает результаты запроса. |
| Save() | Сохраняет данные и параметры текущего выполнения алгоритма. |
| Load() | Читает данные и параметры, сохраненные после выполнения алгоритма. |
| Populations | Текущая популяция запросов. |
| Options | Параметры алгоритма. |
| InitPopulation | Начальная популяция запросов. |
| CurrentPopulation | Текущая популяция запросов. |
| Класс *Population* – свойства популяции запросов и операции, выполняемые над нею. Имеет клоны *InitPopulation* (начальная популяция), *CurrentPopulation* (текущая популяция). | |
| SelectParents() | Выполняет отбор запросов-родителей для формирования пар. |
| CrossingOver() | Выполняет операцию скрещивания пар запросов-родителей. |
| Mutation() | Выполняет операцию мутации запроса с заданной вероятностью. |
| SetQueriesResults | Выполняет запросы популяции и формирует результаты. |
| GetFAttributes() | Возвращает значения промежуточных параметров при расчете целевой функции. |
| Individuals | Запросы (особи) популяции. |
| Resources | Дескрипторы документов, найденных запросами популяции. |
| SigmaFitness | Среднеквадратичное отклонение значений целевой функции запросов текущей популяции. |
| Fitness | Значение целевой функции для популяции (пригодность популяции). |
| LoopNumber | Текущее число проходов алгоритма (число популяций). |
| GenerationNumber | Порядковый номер популяции. |
| SelectedParentsPairs | Родительские пары для последующего скрещивания. |
| Children | Запросы-потомки, полученные после скрещивания пар запросов-родителей. |
| MutedChildren | Запросы-потомки, полученные после мутаций запросов-родителей. |
| CrossingOverTypes | Типы скрещивания |
| Класс *Individual* - свойства запроса (особи) из популяции и операции, выполняемые над ним. Имеет клоны *SelectedParents* (запросы-родители), *SelectedParentsPairs* (пары запросов-родителей), *Children* (запросы-потомки), *MutedChildren* (мутировавшие запросы). | |
| QueryText | Текст запроса. |
| SearchResults | Результаты выполнения запроса. |
| Fitness | Значение целевой функции для запроса (пригодность запроса). |

| Классы и их члены | Описание |
|---|---|
| **Классы и их члены** | **Описание** |
| Класс *Resource* - свойства ресурса, найденного запросом. | |
| Location | Адрес ресурса. |
| Title | Заголовок ресурса. |
| Content | Описание ресурса. |
| TrueContent | Лемматизированное описание ресурса. |
| QueryNumber | Порядковый номер запроса в популяции. |
| Rank | Позиция ресурса в результатах запроса. |
| FitnessAttributes | Результаты расчетов промежуточных параметров целевой функции. |
| Класс *SearchResult* – свойства элемента результатов запроса. | |
| Location | Адрес ресурса. |
| Title | Заголовок ресурса. |
| Content | Описание ресурса. |
| Engine | Поисковая система, с помощью которой найден ресурс. |

Отметим, что в настоящее время для различных программных платформ разработано довольно большое количество реализаций генетических алгоритмов. Причем есть как достаточно старые (но совсем не потерявшие своей актуальности) разработки [4], так и современные реализации, например, GeneHunter (http://www.neuroproject.ru/aboutproduct.php?info=ghinfo) или Genetic Algorithm Framework for .Net (http://johnnewcombe.net/gaf). С другой стороны, некоторые авторы делают попытки унификации и стандартизации подходов к разработкам [5], а другие ориентируются на специальное применение генетических алгоритмов [6, 7].

При разработке основной библиотеки классов обсуждаемого в статье алгоритма GAF авторами было принято решение об оригинальной реализации основных генетических операций и сопутствующих процедур алгоритма. Причины следующие:

1. Особенности вычисления значений целевой функции, которые предусматривают выполнение поисковых операций средствами какой-либо поисковой системы и последующую групповую обработку результатов поиска.

2. Представление поисковых запросов как хромосом, состоящих из генов (ключевых слов), значения которых выражены в номинальной шкале.

3. Принятые специфические интерпретации базовых генетических операций – обмен понятийными элементами поисковых запросов и использование синонимии.

4. Необходимость исследований, анализа и выполнения потенциальных улучшающих модификаций алгоритма.

5. Планируемое использование разработанного генетического алгоритма в качестве основы интеллектуального поискового приложения, работающего на платформе для мобильных устройств.

Указанные причины не позволяют использовать только имеющиеся решения при реализации генетических алгоритмов.

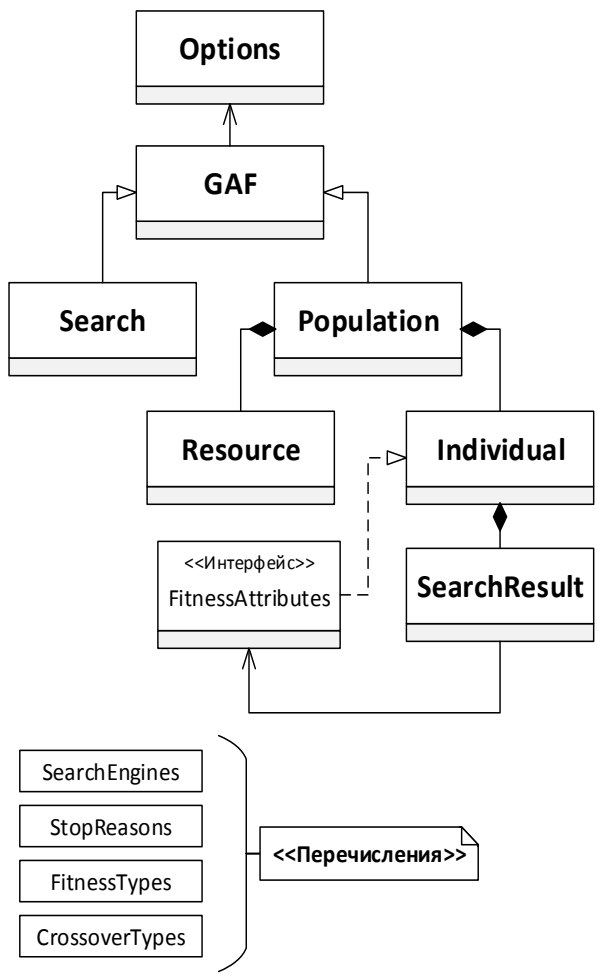
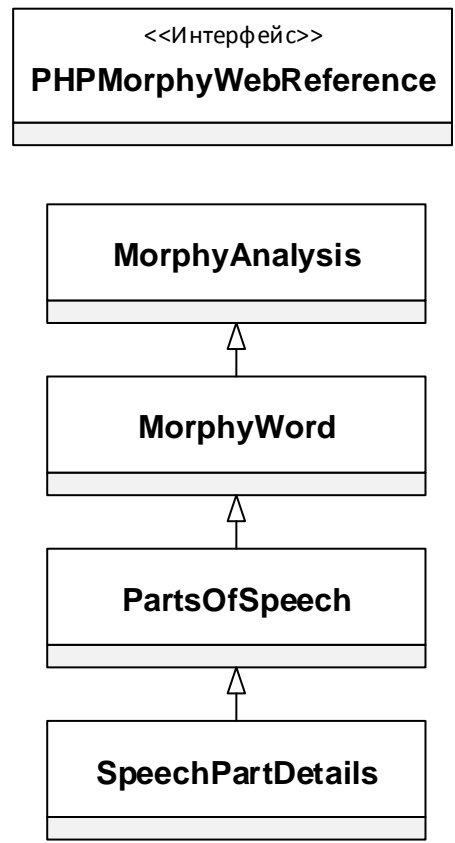

Рис. 3. Объектная модель основной библиотеки классов алгоритма GAF

Рис. 4. Объектная модель программной оболочки MorphyService

**Модуль морфологического анализа и лемматизации запросов**

В этом модуле для первоначальной обработки слов поисковых запросов и результатов поиска (сниппетов и текстовых документов) используется библиотека phpMorphy с бинарными русским и английским словарями (http://phpmorphy.sourceforge.net). Использование phpMorphy описано в http://phpmorphy.sourceforge.net/dokuwiki/manual.

При реализации алгоритма была применена программная оболочка, которая в пространстве имен MorphyService реализует классы, описанные вместе с их основными членами в табл. 2. На рис. 4 приведено графическое изображение укрупненной объектной модели MorphyService.

Таблица 2

| Классы и их члены | Описание |
|---|---|
| Класс *MorphyAnalysis* – коллекция дескрипторов слов для анализа. | |
| MorphyWord [] | Коллекция дескрипторов слов для анализа. |
| Класс *MorphyWord* - коллекция дескрипторов частей речи, соответствующих слову. | |
| partsOfSpeech [] | Коллекция дескрипторов частей речи, соответствующих слову. |
| Класс *PartsOfSpeech* - описание части речи для соответствующего слова. | |
| value | Слово. |
| lemma | Лемма. |
| allForms | Все формы слова. |

| Классы и их члены | Описание |
|---|---|
| SpeechPartDetails [] | Коллекция детальных дескрипторов частей речи. |
| Класс *SpeechPartDetails* – детальное описание части речи для соответствующего слова. | |
| word | Слово. |
| grammems | Граммемы слова |
| Класс *PHPMorphyWebReference* – интерфейс для доступа к модулям морфологического анализа через web-сервис (использован протокол SOAP). | |
| url | Адрес web-сервиса. |
| useCredentials | Параметры разграничения доступа к web-сервису. |
| getMorphyDataAsync() | Выполняет морфологический анализ в асинхронном режиме. |
| getMorphyData() | Выполняет морфологический анализ. |

**Модуль семантического анализа сходства текстов**

Модуль обеспечивает вычисление количественного значения семантической близости $s(d_0, d_i)$, найденного документа $d_i$ (или, по крайней мере, его заголовка и сниппета) и квазидокумента $d_0$ (множества ключевых слов), формирующего поисковый образ документов заданной тематики. Отметим, что в качестве альтернативы $d_0$ могут быть использованы эталонные тексты, адаптивно формируемые в ходе выполнения алгоритма.

Модуль использует экспериментальную платформу, реализующую модель векторного пространства документов [8]. Для доступа к указанной платформе используются объекты класса *TextSimilarity* (см. табл. 3).

Для вычисления $s$ использована модификация модели, в которой каждый документ интерпретируется как вектор $\overline{v}(d) = (w_{1,d}, w_{2,d}, …, w_{M,d})$, где $w_{t,d} = tf_{t,d} * idf_{t,d}$. Здесь $tf_{t,d}$ - частота использования термина в документе, $idf_{t,d}$ - величина, обратная числу документов массива, содержащих данный термин, $idf_{t,d} = \log \frac{P+1}{P_t}$ где $P$ - общее число документов, найденных при выполнении запроса (документы в SERP), $P_t$ - число документов, содержащих данный термин, $M$ - число терминов в $P$ документах. Близость текстов $s$ интерпретирована как косинусная мера близости $s(d_1, d_1) = \frac{\overline{v}(d_1)\overline{v}(d_2)}{\|\overline{v}(d_1)\| \cdot \|\overline{v}(d_2)\|}$, где в числителе скалярное произведение векторов документов $\overline{v}(d_1)$ и $\overline{v}(d_2)$, а в знаменателе – произведение евклидовых норм этих векторов.

Таблица 3

| Классы и их члены | Описание |
|---|---|
| Класс *TextSimilarity* – интерфейс для вычисления сходства двух текстов. | |
| Similarity() | Возвращает значение меры сходства двух текстов. |
| TrueText1 | Текст документа $d_0$ после очистки и морфологического анализа. |
| TrueText2 | Текст документа $d_i$ после очистки и морфологического анализа. |
| SimilarityOptions | Параметры выполнения сравнения документов. |

**Модули поиска**

Модули поиска для Bing и Google реализованы на базе Bing Search API (http://datamarket.azure.com/dataset/bing/search) и Google Custom Search API

(https://developers.google.com/custom-search/json-api/v1/overview) соответственно. Модуль поиска в базе данных XML-документов реализован на базе W3C XML Query (http://www.w3.org/XML/Query).

Все модули выполняют поисковые запросы к соответствующим хранилищам данных и обеспечивают унифицированное представление коллекций результатов. Для активизации модулей поиска из методов класса *Search* (см. табл. 4) разработан следующий общий программный интерфейс (для C#):

*public List<SearchResult> GoogleSearch(string SearchExpression, int ResultNumber),* где

*List<SearchResult>* - типизированный список объектов класса *rdomGAF.SearchResult* (см. описание классов основной библиотеки классов алгоритма);

*SearchExpression* - текст запроса (строка);

*ResultNumber* - количество возвращаемых результатов в списке *List<SearchResult>*.

Отметим, что Bing Search API и Google Custom Search API должны быть лицензированы для коммерческого применения и требуют регистрации приложений.

Таблица 4

| Классы и их члены | Описание |
|---|---|
| Класс *Search* – интерфейсы для работы с поисковыми системами. | |
| *GoogleSearch()* | Выполняет запрос в Google и возвращает результаты. |
| *BingSearch(* | Выполняет запрос в Bing и возвращает результаты. |
| *XMLSearch()* | Выполняет запрос к базе данных XML-документов и возвращает результаты. |

**Модуль управления базой данных**

Модуль обеспечивает доступ к используемым словарям, служебным таблицам, а также сохранение результатов работы алгоритма в реляционной базе данных. Реализует слой доступа к данным общей архитектуры программного обеспечения [9].

Для обеспечения операций манипулирования данными используются объекты класса *SqlHelper* (см. табл. 5). При реализации класса *SqlHelper* использовались компоненты пространства имен *System.Data.SqlClient*, которое является поставщиком данных платформы .Net Framework (набором классов) для доступа к базам данных SQL Server и включает службы доступа к данным ADO.NET.

Перечень основных таблиц используемой базы данных с их описанием представлен в табл. 6.

Таблица 5

| Классы и их члены | Описание |
|---|---|
| Класс *SqlHelper* – интерфейс для операций доступа к базе данных. | |
| OpenConnection() | Открывает соединение с базой данных. |
| CloseConnection() | Закрывает соединение с базой данных. |
| ExecuteCommand() | Выполняет команду SQL или сохраненную процедуру с параметрами. |
| Execute() | Выполняет команду SQL. |

Таблица 6

| Таблица базы данных | Описание |
|---|---|
| edict_terms | Термы. |
| edict_terms_nformes | Нормальные формы термов (леммы). |
| edict_antonyms | Антонимы термов. |
| edict_synonyms | Синонимы термов. |
| twords_documents | Документы (в т.ч. запросы и результаты их выполнения). |
| twords_documents_keywords | Ключевые слова документов. |
| twords_documents_similarity | Данные о семантическом сходстве документов. |
| twords_keywords | Ключевые слова. |

Для обеспечения операций сохранение результатов работы алгоритма используется структура данных, общий вид XML-схемы которой представлен на рис. 5.

```xml
<?xml version="1.0" encoding="utf-8"?>
<xs:schema
  xmlns:xsi="http://www.w3.org/2001/XMLSchema-instance"
  xmlns:xsd="http://www.w3.org/2001/XMLSchema"
  xmlns:xs="http://www.w3.org/2001/XMLSchema"
  attributeFormDefault="unqualified"
  elementFormDefault="qualified">
  <xsd:element name="GAF">
    <xsd:complexType>
      <xsd:sequence>
        <xsd:element name="ErrorText" />
        <xsd:element name="sqlhelper">...</xsd:element>
        <xsd:element name="KeyWords">...</xsd:element>
        <xsd:element name="KeyWordsGenerated" />
        <xsd:element name="Populations">...</xsd:element>
        <xsd:element name="Options">...</xsd:element>
        <xsd:element name="InitPopulation">...</xsd:element>
        <xsd:element name="CurrentPopulati">...</xsd:element>
        <xsd:element name="StopReason" type="xsd:string" />
      </xsd:sequence>
    </xsd:complexType>
  </xsd:element>
</xs:schema>
```

Рис. 5. XML-схема для хранения результатов работы алгоритма *GAF*

При реализации модуля был применен класс *System.Xml.Serialization.XmlSerializer* платформы *.Net* Framework с базовыми методами *XmlSerializer.Serialize()* и *XmlSerializer.Deserialize()*.

**Модуль управления метаданными**

Параметры алгоритма, поддерживаемые модулем управления метаданными, организованы на базе XML-схем. Функциональные группы параметров показаны ниже.

*Основные параметры:*

$g_1$ – используемая поисковая система.

$g_2$ - количество запросов в каждой из генерируемых алгоритмом популяций запросов.

$g_3$ - количество ключевых слов в каждом генерируемом алгоритмом запросе.

$g_4$ - исходный набор ключевых слов и понятий, используемый для генерации запросов.

*Параметры для расчета значения целевой функции:*

$f_1$ - количество результатов поиска, возвращаемых после выполнения запроса.

$f_2$ - количество результатов поиска, возвращаемых после выполнения запросов популяции.

$f_3$ - количество результатов поиска, возвращаемых после выполнения запросов всех популяций.

$f_4$ - коэффициент, учитывающий расположение найденных документов на одном сервере.

$f_5$ - весовой коэффициент для аргумента *p* целевой функции при расчете ранга результата поиска.

$f_6$ - весовой коэффициент для аргумента *r* целевой функции при расчете ранга результата поиска.

$f_7$ - весовой коэффициент для аргумента *s* целевой функции при расчете ранга результата поиска.

$f_8$ - способ вычисления целевой функции для групп результатов поиска при выполнении отдельных запросов или для популяции запросов.

*Параметры генетических операций:*

$c_1$ - множитель для вычисления критерия отбора запросов-родителей при скрещивании.

$m_1$ - Вероятность мутации запроса.

*Параметры завершения алгоритма:*

$e_1$ - заданное число проходов алгоритма.

$e_2$ - предельное значение для среднеквадратичного отклонения целевой функции $\sigma$.

$e_3$ - предельное число проходов алгоритма

*Служебные параметры*: индикатор автоматического сохранения результатов работы алгоритма, имена файлов для сохранения результатов и параметров, строка для связи с базой данных и т.п.

Для доступа к метаданным алгоритма используются объекты класса *Options* (см. табл. 7).

Таблица 7

| Классы и их члены | Описание |
|---|---|
| Класс *Options* – интерфейс для доступа к параметрам алгоритма. | |
| Save() | Сохраняет параметры алгоритма. |
| Load() | Читает сохраненные параметры алгоритма. |
| OptionsCrossover | Параметры операции скрещивания |
| OptionsGeneral | Общие параметры алгоритма |
| OptionsMutation | Параметры операции мутации |
| OptionsFitnessFunction | Параметры целевой функции |
| OptionsStop | Параметры завершения алгоритма |

**Пользовательский интерфейс**

Пользовательский интерфейс (UI) реализован на базе Windows Forms - интерфейса программирования приложений .Net Framework для организации их взаимодействия с пользователями. Интерфейс обеспечивает рабочий цикл алгоритма GAF и показывает ход выполнения отдельных его шагов. Соответственно обеспечивается слой представления (то есть, компоненты UI и логика представления) согласно [9].

На рис. 6 представлена основная панель приложения, реализующего обсуждаемый генетический алгоритм GAF. На рис. 7 представлен фрагмент панели для задания параметров алгоритма GAF.

**Заключение**

Генетический алгоритм, программная реализация которого описана в настоящей статье, является одним из элементов программного обеспечения разрабатываемой интеллектуальной системы информационной поддержки инноваций в науке и образовании [9, 10]. Он играет важную роль в обеспечении адаптивности функционирования поисковых механизмов - повышает эффективность тематического поиска документов за счет повышения качества поисковых запросов и точности оценки релевантности результатов поиска.

Разработанное программное обеспечение алгоритма создает достаточно широкий базис для дальнейших исследований и разработок. Основными направлениями могут быть:

1. Исследование поведения алгоритма на представительной номенклатуре запросов и их тематики. В частности, интерес могут представить зависимости значений целевой функции от количества запросов популяции, количества ключевых слов в запросе, количества проходов алгоритма и других параметров.

2. Уточнение целевой функции. Одно из предложений – введение аргумента, учитывающего условия, задаваемые поисковому агенту параметрами уже найденных документов.

3. Развитие интерпретаций эволюционных операций (кроссинговера и мутации) в контексте создания эффективных поисковых запросов. Так, представляет интерес выявление новых значимых ключевых понятий в процессе обработки результатов поиска для предотвращения преждевременной сходимости.

4. Совершенствование реализации алгоритма, особенно в части приемлемой скорости вычислений.

5. Разработка web-сервиса для обеспечения публичного использования рассмотренной реализации генетического алгоритма как дополнения поисковых систем. Также, по всей видимости, будет интересна мобильная версия пользовательского интерфейса.

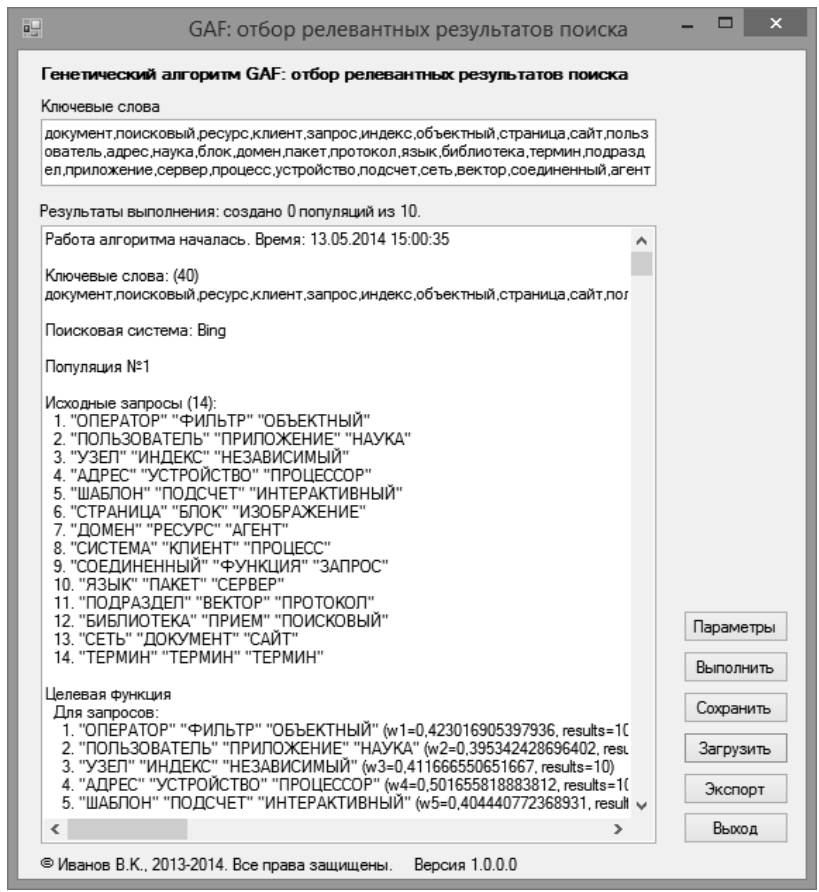

Рис. 6 Основная панель приложения, реализующего генетический алгоритм.

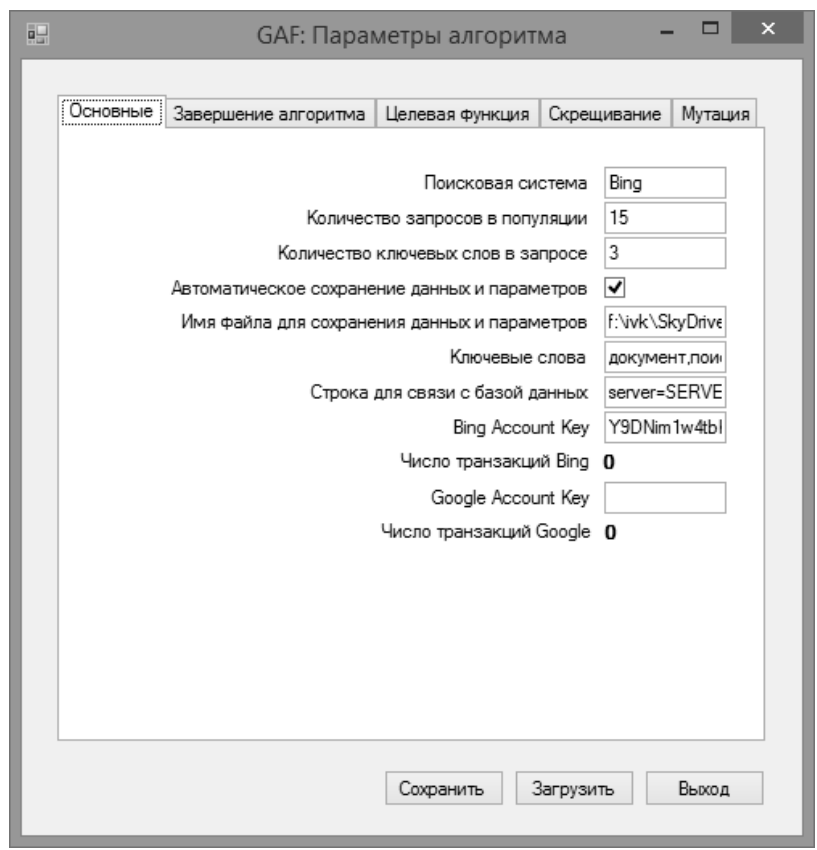

Рис. 7. Панель для задания параметров генетического алгоритма.